\documentclass[12pt]{article}
\usepackage{amsmath,amsfonts,graphics}
\usepackage[dvips]{graphicx}
\providecommand{\abs}[1]{\lvert#1\rvert}
\newcommand{\tr}{\operatorname{tr}}
\newcommand{\re}{\operatorname{Re}}

\newcommand{\hc}{\text{ h.c.}}

\newcommand{\third}{\frac{1}{3}}

\newcommand{\LL}{\mathcal{L}}

\newcommand{\app}{\alpha_{\text{\tiny{++}}}}
\newcommand{\amm}{\alpha_{\text{\tiny{-- --}}}}
\newcommand{\amp}{\alpha_{\text{\tiny{--\! +}}}}
\newcommand{\apm}{\alpha_{\text{\tiny{+\! --}}}}
\newcommand{\Z}{\mathbb{Z}}

\newcommand{\bb}{\mathbf{b}}
\newcommand{\ba}{\mathbf{a}}
\newcommand{\bu}{\mathbf{u}}
\newcommand{\bv}{\mathbf{v}}
\newcommand{\bo}{\mathbf{0}}
\newcommand{\drawsquare}[2]{\hbox{%
\rule{#2pt}{#1pt}\hskip-#2pt
\rule{#1pt}{#2pt}\hskip-#1pt
\rule[#1pt]{#1pt}{#2pt}}\rule[#1pt]{#2pt}{#2pt}\hskip-#2pt
\rule{#2pt}{#1pt}}

\newcommand{\boxnew}{\raisebox{-.5pt}{\drawsquare{7.5}{0.3}}}

\numberwithin{equation}{section}
\begin{document}

\begin{titlepage}
\begin{flushright}
HUTP-01A071\\
BUHEP-02-12\\
\end{flushright}
\vskip 2cm
\begin{center}
{\large\bf Phenomenology of Electroweak Symmetry Breaking from Theory Space}
\vskip 1cm
{\normalsize
\mbox{
\hspace{-0.3in}
Nima Arkani-Hamed$^{a,b}$, Andrew G. Cohen$^{a,c}$, Thomas
Gregoire$^{b}$ and Jay G. Wacker$^{b}$}\\ 
\vskip 0.5cm

(a) Jefferson Physical Laboratory\\
Harvard University\\
Cambridge, MA 02138\\
\vskip .3cm

(b) Department of Physics, University of California\\
Berkeley, CA~~94720, USA\\
\vskip .3cm

(c) Physics Department\\
Boston University\\
Boston, MA 02215\\
\vskip .1in}
\end{center}

\vskip .5cm

\begin{abstract}
  Recently, a new class of realistic models for electroweak symmetry
  breaking have been constructed, without supersymmetry. These
  theories have naturally light Higgs bosons and perturbative new
  physics at the TeV scale.  We describe these models in detail, and
  show that electroweak symmetry breaking can be triggered by a large
  top quark Yukawa coupling.  A rich spectrum of particles is
  predicted, with a pair of light Higgs doublets accompanied by
  new light weak triplet and singlet scalars.  The lightest of
  these new scalars is charged under a geometric discrete symmetry and
  is therefore stable, providing a new candidate for WIMP dark matter.
  At TeV energies, a plethora of new heavy scalars, gauge bosons and
  fermions are revealed, with distinctive quantum numbers and decay
  modes.
\end{abstract}
\end{titlepage}

\section{Introduction}
\label{sec:introduction}
In this decade, experiments will begin to thoroughly explore the
origin of electroweak symmetry breaking. The description of
electroweak symmetry breaking in the Standard Model, in terms of a
fundamental scalar Higgs field, is almost certainly incomplete. The
quadratically divergent radiative corrections to the Higgs mass
suggest the existence of new physics at TeV energies that stabilizes
the weak scale. Until recently, all theories for this stabilization
could be grouped into two categories: those that rely on new strong
dynamics or compositeness near a TeV (such as technicolor, composite
Higgs, or theories with a low fundamental Planck scale), and those
with low-scale supersymmetry. Theories with low scale supersymmetry
are particularly attractive since they allow a perturbative
description of the physics that softens the quadratic divergences, and
naturally lead to light Higgs bosons, as seems favored by precision
electroweak data.

Recently, a qualitatively new category of realistic theories of
electroweak symmetry breaking has been introduced
\cite{Arkani-Hamed:2001nc}. These models, based on the physics of
``theory space''~\cite{Arkani-Hamed:2001ca} (see also \cite{Hill:2000mu}), offer a new mechanism for
softening the quadratic divergences in the Higgs mass.  Electroweak
symmetry breaking is accomplished with naturally light Higgs bosons,
which descend from non-linear sigma model fields whose mass is
protected by ``chiral'' symmetries of the sigma model
(The first attempts at models of this
kind were the ``composite Higgs'' theories
\cite{Kaplan:1984fs,Kaplan:1984sm}).  The physics is perturbative at
energies parametrically above the TeV scale, ultimately requiring UV
completion near $\sim 10 \to 100$ TeV where the non-linear sigma model
fields become strongly coupled. However, the physics of electroweak
symmetry breaking and the new physics at the TeV scale is weakly
coupled. These models are fully realistic, incorporating fermion
masses without producing dangerous flavor-changing neutral currents.
Additionally, there are new stable weakly interacting scalars which
can provide WIMP dark matter.

In \cite{Arkani-Hamed:2001nc} the construction of naturally light
scalars with gauge, Yukawa and quartic self-couplings from theory
space was described, and one fairly simple model of electroweak
symmetry breaking was presented. In this paper, we describe this model
in more detail and begin an exploration of its phenomenology.

We start by describing the bosonic field content and symmetries of
these models which are associated with an $N\times N$ toroidal theory
space (or ``moose'' diagram) and discuss how the light (or ``little'')
Higgs fields emerge at low energies. Subsequently we incorporate the
Standard Model fermions along with their Yukawa couplings. The chiral
symmetry structure of the theory guarantees the absence of quadratic
divergences in the little Higgs masses at low orders in perturbation
theory. We then explicitly compute the leading one-loop radiative
corrections to the little Higgs potential, verifying the absence of
quadratic divergences. We find a calculable negative contribution to
the little Higgs mass squared from the large top Yukawa coupling,
which can naturally drive electroweak symmetry breaking (EWSB).
Finally, we discuss the $N=2$ model in detail, presenting the spectrum
of the theory and the principal decay modes of the new states.

\subsection{Fields and gauge symmetries}
\label{sec:fields}
At energies beneath the scale $\Lambda \sim$ 10--100 TeV, our theory is
well-described by a gauged non-linear sigma model.  The gauge sector
includes $N\times N$ gauge groups $G_{\ba}$, where $\ba = (m,n)$ for
$-N/2 < m,n\le N/2$ and we periodically identify $(m,n) = (m+N,n) =
(m,n+N)$. We take $G_{\ba} =SU(3)$ except for $G_{\bf 0} = SU(2)
\times U(1)$, where ${\bf 0} = (0,0)$.
\begin{figure}[t]
  \centering
  \includegraphics[width=13cm]{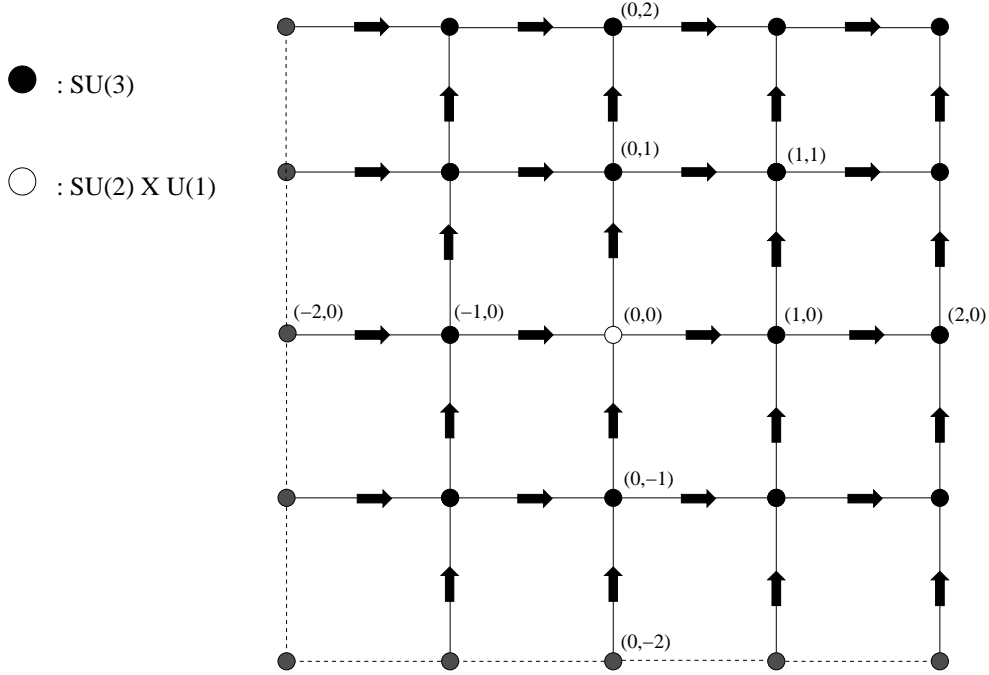}
  \caption{Theory space for the 4 $\times$ 4 torus. Sites $(-2,m)$ and
    $(2,m)$ are identified as are $(n,-2)$ and $(n,2)$. The thick
    arrows represent the link fields.}
  \label{fig:N4moose}
\end{figure}
The non-linear sigma model fields are $3\times 3$ unitary matrices
transforming as bi-fundamentals under nearest neighbor gauge groups.
The gauge group and field content of this model can be represented by
a theory space, shown in Fig. \ref{fig:N4moose} for the case
$N=4$. Each field can be labeled by a ``link'' $l = (\ba,\bb)$, where
$\ba \text{ and } \bb$ are adjacent sites:
\begin{equation}
  \label{eq:1}
  \Sigma_l = e^{i \pi_l }, \quad \text{with} \quad
  \Sigma_{\bar{l}} = \Sigma^{\dagger}_l 
\end{equation}
We adopt the convention $\overline{(\ba,\bb)} \equiv (\bb,\ba)$.  Under
the gauge symmetries $\Sigma_l$ corresponding to link $l=(\ba, \bb)$
transforms as
\begin{equation}
  \label{eq:2}
  \Sigma_l \to g_{\ba} \Sigma_l g^{-1}_{\bb}  
\end{equation}
Here, $g_{\bf 0} = h_{SU(2)} \exp(i \frac{\theta}{\sqrt{3}} T_8)$ is an element of the gauge
group $G_{\bf 0}$, where $h_{SU(2)}$ commutes with $T_8$.
The kinetic part of the action, including gauge interactions, is
\begin{equation}
  \label{eq:3}
  \LL_{K} = - \sum_{\ba} \frac{1}{2 g_{\ba}^2} \tr  F^2_{\ba} +
  \sum_{l} \frac{f_l^2}{4} \tr \abs{D^\mu \Sigma_{l}}^2 + \cdots .
\end{equation}
with a covariant derivative
\begin{equation}
  \label{eq:4}
  D^\mu \Sigma_l = \partial^\mu \Sigma_l +i A_{\ba}^\mu \Sigma_l
  - i \Sigma_l A_{\bb}^\mu .
\end{equation}
For simplicity, we take all the in principle independent $f_l$
constants to be the same. The ellipses in equation \eqref{eq:3}
represent operators involving more than two derivatives suppressed by
powers of the scale $\Lambda$.  The natural size for this cutoff is
$\Lambda = 4\pi f$, where the theory becomes strongly coupled.  Since
we will only be interested in energies well below $\Lambda$ we will
ignore these higher derivative operators.  We also have the usual QCD
$SU(3)_{\text{color}}$ gauge symmetry, under which all these fields
are neutral.

In addition to the operators in \eqref{eq:3} we can include
non-derivative terms as well. We must first classify all operators and
the natural size of their coefficients \cite{Weinberg:1979kz,
  Manohar:1984md}.  Non-derivative gauge invariant operators can be
constructed from ``Wilson'' lines. A line $\ell$ is a collection of
sequential links $l_1,\dots,l_n$ where $l_i = (\ba_i,\ba_{i+1})$.  The
associated Wilson line $W_{\ell}$ is the product of the link fields
$W_{\ell} = \Sigma_{l_1} \Sigma_{l_2} \cdots \Sigma_{l_n}$. Wilson
loops are Wilson lines for closed paths.  We will represent Wilson
loops by pictures of the path with the initial site $\ba$ labeled,
traversed in the counterclockwise direction.  Especially important
will be the fundamental ``plaquettes'', the Wilson loops associated
with the smallest square paths. For example, the symbol ${}_\ba \boxnew$
represents the operator
\begin{equation}
  \label{eq:5}
  {}_\ba \boxnew = \Sigma_{(\ba, \ba + \bu)} \Sigma_{(\ba + \bu, \ba +
    \bu   + \bv)}  
  \Sigma_{(\ba + \bu + \bv, \ba + \bv)} \Sigma_{(\ba + \bv,\ba)} .
\end{equation}
where $\bu,\bv$ are the unit vectors on the moose $\bu = (1,0)$, $\bv
= (0,1)$.

In classifying the natural sizes for the coefficients of operators
constructed from Wilson loops we need to understand their
transformation properties under the chiral symmetries carried by each
link:
\begin{equation}
  \label{eq:16}
  \Sigma_l \to L_l \Sigma_l R_l^\dagger \ .
\end{equation}
A Wilson line breaks the independent $R_{l_i}, L_{l_{i+1}}$ symmetry
transformation to the diagonal $R_{l_i} = L_{l_{i+1}}$, for each $l_i$
in the line. Note that the gauge couplings in \eqref{eq:3} also break
these chiral symmetries, and we expect that gauge interactions will
then induce Wilson loop operators. But these gauge interactions must
come in pairs: any breaking associated with the link $\Sigma_\ell$
must be accompanied by the conjugate breaking of
$\Sigma^\dagger_\ell$. So the only operators which are induced by the
gauge interactions are products of
\begin{equation}
  \label{eq:17}
  \tr W_\ell \tr W^\dagger_\ell \ .
\end{equation}
Since each chiral symmetry breaking is accompanied by a gauge
coupling, the natural size for the coefficient of the operator in
\eqref{eq:17} is $16\pi^2 f^4 (g^2/(16 \pi^2))^n$ where $n$ is the
number of links in $\ell$. Therefore our action includes these terms,
with a naturally small coefficient of this size. This power counting
argument shows that this operator will be renormalized with quadratic
divergences appearing only at $n$ loops. For $N>2$, one loop infrared
contributions of order $g^4/16 \pi^2$ will dominate over the
ultraviolet contributions.

We can in addition add interactions that are not generated in this
way, by including operators that break the chiral symmetries in a
different pattern than the breaking by the gauge couplings. The
coefficients of these operators are therefore in principle unrelated
to the gauge couplings.  We choose the largest new symmetry breaking
interactions to correspond to plaquette operators $\tr {}_\ba \boxnew$
with coefficients $\lambda_\ba$:
\begin{equation}
  \label{eq:6}
  \LL_{\text{pl}} = f^4 \sum_{\ba} \tr {}_\ba \boxnew \, \lambda_{\ba}+
  \hc
\end{equation}
Other gauge invariant operators will be induced with naturally smaller
coefficients. For generic $N$, these plaquettes are the gauge
invariant operators involving the smallest number of link fields. For
$N=2,3$, there are gauge invariant operators involving fewer link
fields (the Wilson lines wrapping cycles of the torus). These however
break different chiral symmetries and are only generated at high loop
order.  Just as the breaking of chiral symmetries associated with the
gauge couplings generates other interactions, the plaquettes induce
all non-derivative gauge invariant operators. For instance, consider a
closed loop $\ell$ which is the boundary of a region tiled by some
collection of plaquettes. Then $\tr W_\ell$ is generated with a
coefficient proportional to $16 \pi^2 f^4$ times the product
$\lambda/16 \pi^2$ for all the tiling plaquettes.  The natural size
for a general gauge invariant operator may be worked out by keeping
track of which chiral symmetries the operator breaks just as we did
with the gauge couplings, remembering that the plaquette operator $\tr
{}_\ba \boxnew$ breaks the chiral symmetries associated with all the
corners of ${}_\ba \boxnew$.  Because of the reduced gauge symmetry at
${\bf 0}$, we may include extra gauge invariant plaquette operators in
the action.  There are four such operators since only four
plaquettes border the ``special'' site:
\begin{equation}
  \label{eq:7} 
  \LL_{\text{pl}_8} = f^4 \tr \sqrt{12}\mathbf{T_8} \left(
    {}_{\bo} \boxnew \, \app + \boxnew_\bo \, \apm
        + \boxnew^\bo \,
    \amm + {}^\bo \boxnew \, \amp \right) + \hc  
\end{equation}
We call these special operators ``$T_8$ plaquettes.''

Finally, the Standard Model fermions are charged only under 
$SU(3)_{\text{color}} \times G_{\bf 0}$, with their familiar Standard Model
quantum numbers. We will see that the non-linear sigma model fields
Higgs the full gauge symmetry down to the Standard Model $SU(3)_{\text{color}} 
\times SU(2)_L \times U(1)_Y$ symmetry. 

\subsection{Discrete Symmetries}
\label{sec:discrete-symmetries}

The site, link  and plaquette structure of theory space allows for discrete
symmetries associated with geometric transformations. These include
rotations $R$ by $90^\circ$ about an axis through the site $\bo$
perpendicular to the theory space.
This takes a site $\ba = (m,n)$ into $R\ba = (-n,m)$. On a
link $l=(\ba,\bb)$ $R$ acts as $R_l = (R\ba,R\bb)$. We also have
theory space ``parities'' $P_{u,v}$ which correspond to reflection
about the $\bu$ and $\bv$ axes respectively: $P_u \ba = (-m,n)$, $P_v
\ba = (m,-n)$, and similar action on the links.  Note that these eight
discrete transformations, comprising the point symmetry group $D_{4}$,
can be generated by the two transformations $R$ and $P_u$. On fields
\begin{subequations}
  \label{eq:sub1}
  \begin{alignat}{2}
    R A^\mu_\ba &= A^\mu_{ R\ba}  &\qquad P_u A^\mu_\ba &= A^\mu_{P_u \ba} \\
    R\Sigma_l &= \Sigma_{R l} & \qquad   P_u\Sigma_l &= \Sigma_{P_u l} 
  \end{alignat}
\end{subequations}
The action of these transformations on plaquettes follows from these
rules:
\begin{subequations}
  \begin{align}
    R\, {}_\ba \boxnew &= \boxnew_{R\ba} \\
    P_u\, {}_\ba \boxnew &= \left(\boxnew_{P_u \ba}\right)^\dagger
  \end{align}
\end{subequations}

If we demand that these transformations are symmetries of the action,
the gauge and plaquette couplings must satisfy the relations
\begin{subequations}
  \begin{alignat}{2}
    g_{R\ba} &= g_\ba  &\qquad  g_{P_u \ba} &= g_\ba \\
    \lambda_{R \ba-\bu} & = \lambda_\ba  & \qquad
    \lambda_{P_u\ba-\bu}  &= \lambda^*_\ba 
  \end{alignat}
\end{subequations}
The rotational symmetry forces the four special plaquette couplings to
be equal $\app=\apm=\amp=\app\equiv\alpha+i\epsilon$, while the parity
symmetry forces $\epsilon=0$.  We are not obliged to impose these
symmetries exactly, and in fact, our final Lagrangian will preserve
only the $\Z_4$ subgroup generated by the $90^{\circ}$ rotation $R$.
Other patterns of symmetry breaking are also possible.

\subsection{Little Higgs}

We can now discuss the origin of natural
electroweak symmetry breaking in this class of models. We will later
analyze the specific case $N=2$ in detail.
 
Since the theory (so far) has a $D_4$ symmetry, we take $\epsilon=0$,
keeping only the largest couplings $g_\ba, \lambda_\ba, \text{ and }
\alpha$.  We will assume that these couplings are all perturbative; as
we have discussed all other operators have smaller effects and may be
treated as further perturbations.

In this gauged non-linear sigma model only some subgroup of the gauge
symmetry is realized at low energies---many of the $\pi_l$ are
``eaten'', giving masses to many of the gauge bosons. We will choose
the plaquette couplings such that the vacuum field configuration is
near $\Sigma_l = 1$.  Under the uniform gauge transformation $g_\ba =
g_{SU(2)\times U(1)}$ independent of $\ba$, the configuration
$\Sigma_l = 1$ is invariant, showing that the diagonal $SU(2)_L\times
U(1)_Y$ gauge group is left unbroken.  Thus there
are massless $SU(2)_L\times U(1)_Y$ gauge bosons together with a spectrum
of gauge bosons with mass scale set by $g f$ with $g$ a typical gauge
coupling. In order to see this explicitly and identify the spectrum
of the theory, we expand the Lagrangian to quadratic order in
$A^\mu_{\ba}$ and $\pi_l$. It is convenient to define link fields
pointing in the $\bu,\bv$ directions as
\begin{equation}
  \Sigma_{(\ba, \ba + \bu)} \equiv e^{i \pi^u_\ba},  \quad 
  \Sigma_{(\ba, \ba + \bv)} 
  \equiv e^{i \pi^v_\ba}
\end{equation}
The Lagrangian to quadratic order is 
\begin{equation}
  \begin{split}
    \LL^{\text{quad}} = & - \sum_{\ba} \frac{1}{2 g_{\ba}^2} \tr
    F^2_{\ba} + \sum_{\ba} \tr \frac{f^2}{4} \left(\partial^\mu
      \pi^u_\ba + A^\mu_\ba - A^\mu_{\ba
        + \bu} \right)^2 + u \to v \\
    & + \sum_{\ba} f^4 \lambda_\ba \tr \left(\pi^u_\ba + \pi^v_{\ba +
        \bv} - \pi^u_{\ba + \bv} - \pi^v_{\ba} \right)^2 +
    \LL^{\text{quad}}_{\text{pl}_8}
  \end{split}
\end{equation}
where to save space we have not written the $T_8$ plaquette
contribution explicitly.  From the quadratic piece involving the gauge
bosons we see explicitly that only the uniform configurations with
$A^\mu_\ba = A^i_{SU(2)} {\bf T}^i + A_{U(1)} {\bf T_8}$ are massless,
with all remaining $N^2 - 1$ gauge boson octets acquiring mass. Since
we have $2 N^2$ link scalars $\pi_l$, this means that $2
N^2 - (N^2 - 1) =  N^2 + 1$ scalars remain uneaten.  The
plaquettes give a mass to $N^2 - 1$ independent combinations of these
scalars (there are $N^2$ plaquette terms but the sum of all $N^2$
linear combinations that appear vanishes, so only $N^2 - 1$
independent combinations become massive). Therefore we expect $ (N^2
+ 1) -  (N^2 - 1) =  2$ scalars to be massless at this order.
These zero modes are easy to identify in this case: they are the
uniform configurations for the fields $\pi^u_\ba, \pi^v_\ba$
\begin{equation}
  \pi^u_\ba = \frac{2 u}{N f}, \quad \pi^v_\ba = \frac{2 v}{N f}
\end{equation}
where we have included the factor $2/(N f)$ to canonically normalize
$u$ and $v$.

Since there is an unbroken discrete $\Z_4$ symmetry
under which $u \to v$ and $v \to -u$, it is convenient to group $u,v$
into a field that transforms homogeneously under this $\Z_4$. Defining
${\cal H} = (u + i v)/\sqrt{2}$, we have ${\cal H} \to -i {\cal H}$,
${\cal H}^{\dagger} \to i {\cal H}^\dagger$ under the $\Z_4$ symmetry.
The components of the $3 \times 3$ matrix ${\cal H}$ are
\begin{equation}
  {\cal H} = \frac{u + i v}{\sqrt{2}} \equiv
  \begin{pmatrix} \phi + \eta/2 \sqrt{3} &
    h_1/\sqrt{2} \\ h_2^{\dagger}/\sqrt{2} & -\eta/\sqrt{3} 
  \end{pmatrix}
\end{equation}
where under the Standard Model $SU(2)_L \times U(1)_Y$ gauge symmetry,
$\phi,\eta$ are complex fields transforming respectively in the
$\text{\bf 3}_0, \text{\bf 1}_0$ representation, and $h_1, h_2$ have
the quantum numbers $\text{\bf 2}_{1/2}$ of the Standard Model Higgs.

Going beyond quadratic order these zero modes have gauge
interactions under the unbroken $SU(2)_L \times U(1)_Y$ gauge symmetry, as
well as a quartic coupling which can be found by expanding the plaquette
action and integrating out the heavy modes. In the case where all
the $\lambda_{\ba}$ are equal, in accordance with the extra-dimensional
intuition, there is no heavy-light-light scalar coupling thus we don't need
to integrate out the heavy modes and we find:
\begin{equation}
  \label{eq:quartic}
  \text{constant} + 4 \lambda \tr\abs{[{\cal H},{\cal H}^{\dagger}]}^2
  + \dots  \supset \lambda \tr (h_1 h_1^{\dagger}-h_2 h_2^{\dagger})^2
  + \lambda  (h_1^{\dagger}h_1-h_2^{\dagger}h_2)^2  
\end{equation}
where $\lambda =4/N^2 \re\,\lambda$.\footnote{For general $\lambda_{\ba}$,
the potential is given by $16 \left(\sum_{\ba}(\re\,\lambda_{\ba})^{-1}\right)^{-1} \tr\abs{[{\cal H},{\cal H}^{\dagger}]}^2$}
We have again neglected the $T_8$
plaquettes, which do not affect the qualitative discussion.  Note that
the Higgs interactions are very similar to the quartic potential for
the Higgses of the supersymmetric Standard Model (MSSM). The
commutator form naturally arises from the plaquettes and is essential
for obtaining a tree-level quartic potential but no tree-level mass.
This is crucial for generating a natural hierarchy while incorporating
hard quartic couplings.  It is this feature that distinguishes these
theory space models from previous attempts at obtaining the Higgs as a
pseudo-Goldstone boson.

We found that, classically, we have two massless modes with quartic
interactions. Radiative corrections will produce masses for these
scalars, but as we have argued they are free of quadratic
divergences at one-loop. Thus they are naturally light compared to the
cutoff of the theory. We will call these naturally light scalars
little Higgses.  For a general theory space the spectrum of little
Higgses can be determined from the topology of the theory space, as
described in \cite{moose}. We may introduce masses for the little
Higgses classically by turning on a non-zero value for $\epsilon$.
As mentioned in the beginning of the section, up to now the
theory has a $D_4$ symmetry (under which $(h_1,h_2)$ transform as a
doublet) which enforces $\epsilon =0$.  The fermion
couplings will break this $D_4$ symmetry, allowing us to preserve at
most the $\Z_4$ subgroup. Although not essential, this $\Z_4$ symmetry
has several phenomenological virtues, including helping
prevent $\eta$ from acquiring a vev, and we will keep it as an exact
symmetry of our model.

\subsection{Fermions}
\label{sec:fermions}

The Standard Model fermions have their conventional charges under the
$G_{ \bo} = SU(2) \times U(1)$ symmetry; they therefore have the same
charges under the diagonal subgroup that becomes the Standard Model
$SU(2)_L \times U(1)_Y$. In order for these fermions to acquire mass,
they need Yukawa couplings to the Higgs fields, which must arise from
gauge-invariant couplings to the $\Sigma_l$'s. Since the $\Sigma_l$
are $3 \times 3$ matrices and naturally act on triplets, we  write
the SM fermions as 3 components column vectors. In terms of left
handed Weyl fields:
\begin{equation*}
   Q = \left( \begin{array}{c} q\\0\end{array}\right)
  \qquad
  U^c = \left( \begin{array}{c} 0\\0\\u^c\end{array}\right).
  \qquad D^c = \left(\begin{array}{c} 0 \\0 \\d^c\end{array}\right)
  \qquad L = \left(\begin{array}{c} l\\0\end{array}\right)
  \qquad E^c = \left(\begin{array}{c} 0 \\0 \\e^c\end{array}\right)
\end{equation*}
Yukawa couplings come from gauge-invariant operators of the form
\begin{equation*}
  Q^T W_{\bo \to \bo} U^c
\end{equation*}
where $W_{\bo \to \bo}$ is some linear combination of  Wilson
loops that start and end at $\bo$. A convenient set of such Wilson
loops is provided by ${\cal U} = \Sigma_{\bo ,\bu} \Sigma_{\bu , 2
  \bu} \cdots \Sigma_{-\bu, \bo}$ and ${\cal V} = \Sigma_{\bo ,\bv}
\Sigma_{\bv , 2 \bv} \cdots \Sigma_{-\bv, \bo}$, together with ${\cal
  U}^{\dagger}$ and ${\cal V}^{\dagger}$.  Under the $\Z_4$ discrete
symmetry, ${\cal U} \to {\cal V}, {\cal V} \to {\cal U}^\dagger$. It
is convenient to form linear combinations that transform
homogeneously under the $\Z_4$:
\begin{subequations}
  \begin{align}
    {\cal U} + {\cal U}^\dagger \pm ({\cal V} + {\cal
      V}^\dagger) &\to  
    \pm \left[{\cal U} + {\cal U}^{\dagger} \pm ({\cal V} + {\cal
        V}^{\dagger})\right]  \\
    {\cal U} - {\cal U}^\dagger \pm i ({\cal V} - {\cal
      V}^\dagger)   & \to 
    \mp i 
    \left[{\cal U} - {\cal U}^\dagger \pm i ({\cal V} - {\cal
        V}^\dagger)\right] \\
    \intertext{where expanding to low order in $u$ and $v$}
    {\cal U} + {\cal U}^{\dagger} \pm ({\cal V} + {\cal
      V}^{\dagger}) & \sim  
    \text{const} + {\cal O}\left({\cal H}^2/f^2\right) + \cdots \notag\\
    {\cal U} - {\cal U}^\dagger + i ({\cal V} -
    {\cal V}^\dagger)  \sim i {\cal H}/f + \cdots
     &\qquad {\cal U} - {\cal U}^\dagger - i ({\cal V} -
    {\cal V}^\dagger)  \sim i {\cal H}^{\dagger}/f + \cdots \notag
  \end{align}
\end{subequations}
Since the first combination begins at quadratic order in
the Higgs fields, it does not produce a  Yukawa coupling for the SM
fermions. We therefore use the second combination in the Yukawa coupling:
\begin{equation}
  \label{eq:8}
  \lambda_u f  Q^T \left[{\cal U} - {\cal U}^\dagger \pm i ({\cal V}
    - {\cal V}^\dagger) \right] U^c 
\end{equation}
where the $\Z_4$ is preserved with $q,u^c$ transforming so that $(q
u^c) \to \pm i (q u^c)$. Choosing the plus sign and expanding to
linear order in the Higgs field, we get:
\begin{equation}
  \label{eq:yuk}
  \lambda_u q h_1 u^c
\end{equation}

In some models, like the one we study later in more detail, only $h_1$
gets a vev. In that case, in order to give a non-vanishing mass to
both the up and down type quarks, we choose the opposite sign for the
down sector:
\begin{equation}
  \lambda_d f  {D^c}^T \left[{\cal U} - {\cal U}^\dagger - i ({\cal V}
    - {\cal V}^\dagger) \right] Q = \lambda_d h_1^{\dagger} q d^c + \cdots
\end{equation}
With this choice, $q, d^c$ transforming so that $(q d^c) \rightarrow
-i (q d^c)$ preserves the $\Z_4$ symmetry.  We can write a similar
operator for the leptons:
\begin{equation}
  \lambda_l f E^T \left[{\cal U} - {\cal U}^\dagger - i ({\cal V}
    - {\cal V}^\dagger) \right] L  = \lambda_l h_1^{\dagger}l e + \cdots
\end{equation}

Each of these operators is non-local in theory space and each breaks
\emph{all} the chiral symmetries protecting the little Higgs mass.
Their inclusion directly in the Lagrangian would re-introduce
quadratic divergences at one-loop of order
$\lambda^2\Lambda^2/(4\pi)^2$. This is in contrast to the gauge
couplings and plaquettes which only collectively break the chiral
symmetries, introducing quadratic divergences only at higher order.
The divergences generated by these non-local operators are negligible
for all fields except the top quark. We will therefore generate the
top Yukawa coupling from local interactions in theory space by
effectively spreading the top quark out in theory space. We do this by
introducing massive vector-like fermions on the sites and coupling
them to the chiral fermions on the special site through link fields.
Upon integrating out the massive fermions, the top Yukawa coupling is
generated.  The $\Z_4$ symmetry requires the introduction of two sets
of vector-like fermions, $\chi$'s and $\tilde{\chi}$'s. We place them
along the $\bu$ and $\bv$ axes: $\chi_{p \bu}, \tilde{\chi}_{p \bu}$
at the site $p \bu$, where $p$ is a an integer, and $\chi_{p \bv},
\tilde{\chi}_{p \bv}$ at the site $p \bv$, all of which are triplets
under the $SU(3)$ gauge symmetry at the corresponding site. In
addition, each of these fermions are triplets under $SU(3)$ color.
Finally they also carry charge $ 1/3$ under the $U(1)$ gauge factor in
$G_{\bf 0}$. We also introduce the conjugate of each of these fields:
\begin{equation}
  \begin{aligned}
    \chi_{p \bu}, \tilde{\chi}_{p \bu} & \sim \left( \mathbf{3}_c\times
      \mathbf{\bar{3}}\right)_{\mathbf{+\third}}
    & \qquad
    \chi_{p \bu}^c, \tilde{\chi}_{p \bu}^c  & \sim  \left(
      \mathbf{\bar{3}}_c\times \mathbf{3}\right)_{\mathbf{-\third}}\\ 
    \chi_{p \bv}, \tilde{\chi}_{p \bv} & \sim \left( \mathbf{3}_c\times
      \mathbf{\bar{3}}\right)_{\mathbf{+\third}}
    & \qquad
    \chi_{p \bv}^c, \tilde{\chi}_{p \bv}^c  & \sim  \left(
      \mathbf{\bar{3}}_c\times  \mathbf{3}\right)_{\mathbf{-\third}}
  \end{aligned}
\end{equation}

\begin{figure}[t]
  \centering
  \includegraphics[width=12cm]{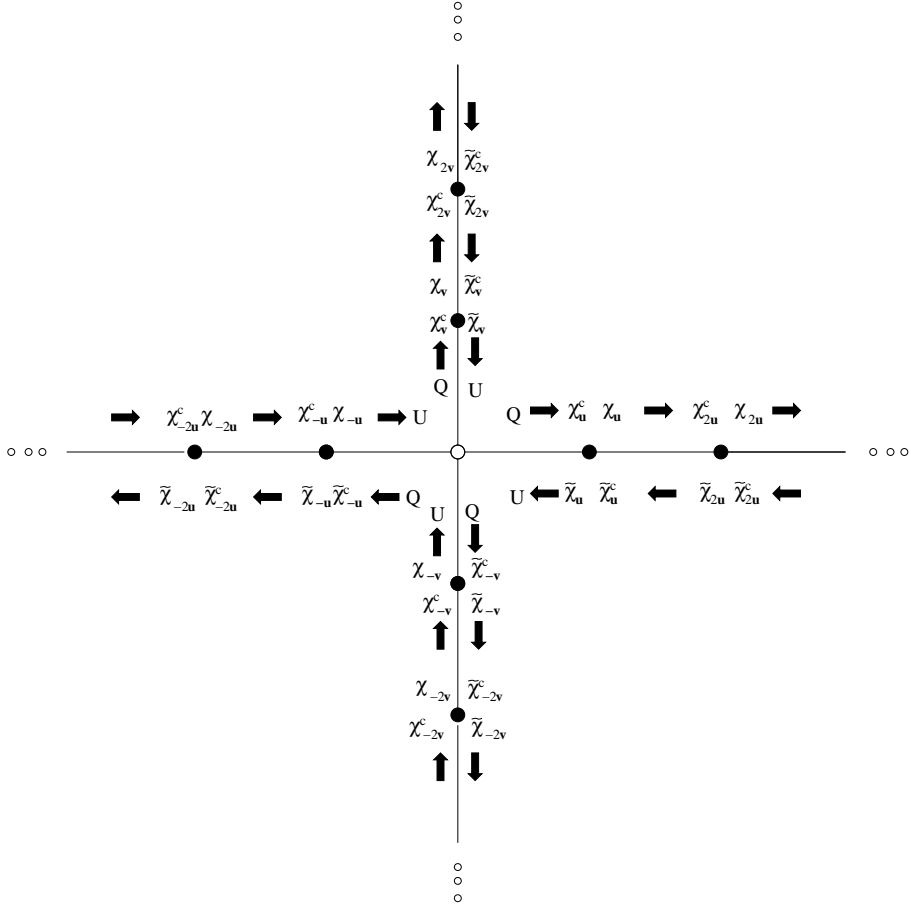}
  \caption{Representation of top quark Lagrangian}
  \label{fig:fermion}
\end{figure}

The interactions can be represented by the ``traffic pattern'' of
Fig. \eqref{fig:fermion}. Note that the $\Z_4$ symmetry is manifest
in this figure and will be preserved with the following charge
assignment for the fermions:
\begin{equation}
  \begin{aligned}
  Q & \rightarrow & e^{i \frac{\pi}{4}}  Q & \rightarrow & i Q &
      \rightarrow & e^{i\frac{3 \pi}{4}} Q & \rightarrow & - Q\\
      U^c & \rightarrow & e^{i \frac{\pi}{4}}  U^c & \rightarrow & i U^c &
      \rightarrow & e^{i\frac{3 \pi}{4}} U^c & \rightarrow & - U^c  \\
       \chi_{p\bu} & \rightarrow & \chi_{p\bv} &
      \rightarrow & \tilde{\chi}_{-p \bu} & \rightarrow & \tilde{\chi}_{-p
        \bv} & \rightarrow & - \chi_{p\bu} \\ 
      \chi_{p\bu}^c & \rightarrow & \chi_{p \bv}^c & \rightarrow &
      \tilde{\chi}_{-p \bu}^c & \rightarrow & \tilde{\chi}_{-p \bv}^c &
      \rightarrow & - \chi_{p \bu}^c 
  \end{aligned}
\end{equation}
The explicit Lagrangian is given by:
\begin{equation}
  \begin{split}
    \mathcal{L}_{\text{top}} &= M_L' Q_3^T\left(\Sigma_{\bo,\bu}
      \chi_{\bu}^c + i \Sigma_{\bo,-\bu} \tilde{\chi}_{-\bu}^c\right)
    +M_R'\left(\chi_{-\bu}\Sigma_{-\bu,\bo} + i \tilde{\chi}_{\bu}
      \Sigma_{\bu,\bo}\right)U_3^c  \\
    &\ + e^{i \frac{\pi}{4}}\left[ M_L'
      Q_3^T\left(\Sigma_{\bo,\bv} \chi_{\bv}^c + i \Sigma_{\bo,-\bv}
        \tilde{\chi}_{-\bv}^c\right)
      +M_R'\left(\chi_{-\bv}\Sigma_{-\bv,\bo} + i \tilde{\chi}_{\bv}
        \Sigma_{\bv,\bo}\right)U_3^c\right] \\ 
    &\ +\sum_{p\neq0} \Big\{\chi_{p \bu}\left(M \chi_{p \bu}^c+M'
      \Sigma_{p \bu,(p+1)\bu} \chi_{(p+1) \bu}^c\right) 
    +  \tilde{\chi}_{-p \bu}\left(M \tilde{\chi}_{-p \bu}^c+M'
      \Sigma_{-p \bu,-(p+1) \bu} \tilde{\chi}_{-(p+1)\bu}^c\right) \\ 
    &\ + e^{i \frac{\pi}{4}} \left[\chi_{p \bv}\left(M \chi_{p
          \bv}^c+ M'\Sigma_{p \bv,(p+1)\bv} \chi_{(p+1) \bv}^c\right)
      + \tilde{\chi}_{-p \bv} 
      \left(M \tilde{\chi}_{-p \bv}^c +M'\Sigma_{-p \bv,-(p+1) \bv}
        \tilde{\chi}_{-(p+1) \bv}^c \right)\right]\Big\} 
  \end{split}
\end{equation}
where, for notational simplicity we have chosen the same $M$ for every
site. Other realistic models with a smaller fermion structure can be
found in \cite{moose}.
Integrating out the massive fermions gives the top quark Yukawa
coupling. Extracting this Yukawa coupling requires the diagonalization
of the full mass matrix, but if all the $M$'s are
comparable, it is given parametrically by $\lambda_t \sim M/f$.  This
method of implementing Yukawa couplings preserves a $U(2)^5$ flavor
symmetry broken only by the the Yukawa matrices of the first two
generations, ensuring the absence of dangerous flavor changing
neutral currents.
 
The Lagrangian 
\begin{equation}
  \LL = \LL_K + \LL_{\text{pl}} + \LL_{\text{pl}_8} +\LL_{\text{fermions}}
  +\LL_{\text{other}} 
\end{equation}
specifies our model, where $\LL_{\text{other}}$ includes higher
dimension operators suppressed by powers of $4\pi f$ as well as
non-derivative operators with naturally sized coefficients.  The
parameters are $g_\ba, \lambda_\ba,\alpha, \epsilon$, and the
dimensionful parameters $f$, $M$'s, $M_L'$ and $M_R'$.  $g_\ba$,
$\lambda_a$ and $\alpha$ are independently renormalized, and we
take them to be perturbative with sizes $\lambda_\ba, \alpha \sim
g_\ba^2$. The fermion masses $M's$,$M_L'$ and $M_R'$ are also
independently renormalized and protected by chiral symmetries. Finally
$\epsilon$, which breaks the $D_4$ symmetry as well as the chiral
symmetries of the $T_8$ plaquette is then renormalized through a
combination of $M_{L,R}'$ and $\alpha$ at two loops, giving a tiny
(and irrelevant) natural size $\epsilon \sim \alpha {M_L'}^2 {M_R'}^2
/f^2 (16\pi^2)^3$.

\subsection{Radiative Corrections}
As we have argued, the chiral symmetries protecting the little Higgs
are only broken by combinations of several couplings, and therefore
quadratic divergences are absent at one loop. In the case $N\geq 3$
the little Higgs potential is dominated by finite IR effects and is
calculable. In the $N=2$ case, there is in general a log divergence at
one-loop and the UV and IR contributions are comparable. While our
spurious symmetry analysis guarantees this result, it is reassuring to
see it arise in perturbative calculation by computing the one-loop
Coleman-Weinberg potential.  We turn on the little Higgs fields $v$,
and calculate the mass matrix $M(v)$ of the theory in presence of
these background fields. The quadratic divergence is proportional to
$\tr M(v)^{\dagger}M(v)$ and the logarithmic divergence to $\tr
(M(v)^{\dagger}M(v))^2$. As an example, we look at the simple circular
theory space studied in \cite{Arkani-Hamed:2001nc} with only 3 sites.
Turning on a uniform background value for the link fields: $U_i = e^{i
  v \sigma_3/f \sqrt{3}}$, the mass matrix squared, $M^{\dagger} M$,
for the charged gauge bosons is proportional to:
\begin{equation}
  \begin{pmatrix}
    2& -e^{-i v/f \sqrt{3}}& -e^{i v/f \sqrt{3}}\\
    -e^{i v/f \sqrt{3}}& 2& -e^{-i v/f \sqrt{3}} \\
    - e^{-i v/f \sqrt{3}} &-e^{iv/f \sqrt{3}}&2
  \end{pmatrix}
\end{equation}
We can easily see that $\tr M^{\dagger} M$ doesn't depend on $v$ since
$v$ doesn't appear anywhere on the diagonal. In terms of mass
eigenstates, turning on v increases the mass squared of light fields
but decreases the mass of heavy fields. It is similarly easy to see
that $\tr \left(M^{\dagger} M\right)^2$ doesn't depend on $v$, so
there is neither quadratic nor log divergence at one loop in this
model. In the $N=2$ case, the mass matrix is proportional to:
\begin{equation}
  \begin{pmatrix}
    2&2 \cos(v/f\sqrt{2}) \\
    2\cos(v/f \sqrt{2})&2 \end{pmatrix}
\end{equation}
Again there is no $v$ on the diagonal and thus no quadratic
divergence. However $\tr\left(M^{\dagger}M\right)^2 =
8\left(1+\cos^2(v/f \sqrt{N})\right)$, and so there is a log
divergence in this case.

In our model with a 2 dimensional lattice of sites the mass matrices
are larger, but the same observations hold true.  Radiative
corrections  generate mass terms:
\begin{equation}
  \label{eq:quadratic}
  V_{\text{quadratic}} = m_1^2 \abs{h_1}^2 + m_2^2
  \abs{h_2}^2+m_{\phi}^2 \tr \abs{\phi}^2+ m_{\eta}^2 \abs{\eta}^2
\end{equation}
Note that there is no $h_1^{\dagger}h_2$ term or linear term in $\eta$
due to the $\Z_4$ symmetry.  If $m_{\phi}^2,m_{\eta}^2,m_2^2>0$ and
$m_1^2<0$, we will trigger electroweak symmetry breaking (EWSB). Since
the quartic potential vanishes for $h_1=h_2$, we also require
$m_1^2+m_2^2>0$ to avoid rolling away along this flat direction. One
might worry that after EWSB, $\phi$ and $\eta$ could get large vevs.
However, the $\Z_4$ symmetry which leads to an unbroken $\Z_2$
symmetry even after electroweak symmetry breaking forbids odd powers
of $\phi$ and $\eta$. With positive $m_{\phi,\eta}^2$, both vevs
vanish.

For $N\geq 3$, the mass terms in \eqref{eq:quadratic} are dominated by
finite contributions which are in principle calculable. In the $N=2$
case, most of them are log divergent at one loop and therefore not
calculable.  However, a spurion analysis shows that even in that case,
the mass \emph{difference} between $h_1$ and $h_2$ is dominated by IR
effects and is finite at one loop. In the limit where Yukawa couplings
are neglected, the theory has a $D_4$ symmetry under which $h_1$ and
$h_2$ form a doublet and are therefore degenerate. Consequently, the
mass difference must come from fermion loops. By computing the
fermionic contribution to the one-loop Coleman-Weinberg potential when
$N=2$, we find that this is indeed the case, the result being:
\begin{equation}
  \label{eq:topCW}
  -\left(\frac{3 M^2 {M_R'}^2 {M_L'}^2}{2 \pi^2 f^2
      ({M_R'}^2-{M_L'}^2)} \log     \frac{M^2+4 {M_R'}^2}{M^2+4
      {M_L'}^2}\right) \abs{h_1}^2 + {\cal O}(h^4) 
\end{equation}
This is negative for any values of $M_L',M_R'$ and therefore can
trigger EWSB in a natural way. This is an improvement over
\cite{Arkani-Hamed:2001nc} where EWSB was triggered by a non-zero
$\epsilon$; here it is driven by top quark radiative corrections, much
as in the MSSM.

\subsection{Scales}
In the very low energy theory, keeping only the classically massless
modes, the Higgs would get a quadratically divergent mass. It is the
appearance of the heavy modes that cancel this divergence. For
example it is the appearance of heavy ``KK'' gauge bosons at energy $g
f/N$ which cancels the divergence of the low energy gauge boson loop.
Similarly, it is the heavy fermions introduced to ``delocalize'' the
top quark which cancel the divergent top quark loop of the very low
energy theory. Finally, the heavy scalars which get masses from the
plaquettes are necessary to cancel the divergence from the Higgs
quartic couplings.  This observation allows us to estimate the
various scales of the theory such as the Higgs mass and the cutoff $4
\pi f$. For example the one loop gauge contribution to the Higgs mass
is approximately
\begin{equation} 
  m_h^2 \sim \frac{g_{LE}^2}{16 \pi^2} \left(\frac{g f}{N}\right)^2
  \sim \left(\frac{g^2_{LE}}{16 \pi^2}\right)^2 \left(4 \pi f\right)^2
  \sim \left(\frac{\alpha_{LE}}{4 \pi}\right)^2 \Lambda^2
\end{equation}
where $g_{LE}^2 = g^2/N^2$ is the low energy coupling, i.e. the
coupling of the unbroken gauge group. This is to be contrasted with
the Standard Model case where:
\begin{equation}
  m_h^2 \sim \left(\frac{\alpha_{LE}}{4 \pi}\right) \Lambda^2
\end{equation}
The presence of an extra $\alpha_{LE}/4 \pi$ factor in our case allows
us to take the cutoff of our theory parametrically above the TeV
scale, near $\sim$ 10--100 TeV, and still have perturbative new physics
at the TeV scale stabilizing the Higgs. A similar conclusion can be
reached for the quartic couplings.

The same simple estimate can be done for the top quark contribution,
but it is more model dependent:
\begin{equation}
  m_h^2 \sim \frac{\lambda_t^2}{16 \pi^2} M^2 \sim
  \left(\frac{\lambda_t^2}{16 \pi^2}\right)^2 \left(4 \pi f\right)^2 \sim
  \left(\frac{\alpha_t}{4 \pi}\right)^2 \Lambda^2
\end{equation}
where $\lambda_t$ is the top Yukawa coupling given parametrically by
$\sim M/f$. Again the additional $\alpha_t/4 \pi$ factor allows us to
choose the cutoff in the $\sim 10$ TeV region.

To summarize, in the very low energy theory beneath the masses of the
heavy modes, we have the Standard Model with 2 Higgs doublets and
other scalars in the $\sim 100$ GeV region. At roughly $1$ TeV, heavy
fermions, scalars and vector bosons responsible for stabilizing the
Higgs mass appear. Finally, at $\sim$ 10--100 TeV our description of
the physics in terms of non-linear sigma model fields breaks down and a
UV completion is needed. Although we can imagine many different
possibilities for this UV completion (such as SUSY, new strong
dynamics, low scale quantum gravity, etc.), they do not affect our
description of electroweak symmetry breaking and we do not discuss it
here.  The important point is that the cutoff of our theory (where it
becomes strongly coupled and a UV completion is needed) is
parametrically above the scale where perturbative new physics
appears to stabilize the weak hierarchy.

\subsection{Dark Matter}
\label{sec:dark}
As we have seen, our theory has a natural geometric $\Z_4$ symmetry,
generated by the $90^\circ$ rotation $R$ on theory space. The scalars
$\phi,\eta,h_1,h_2$, together with the Standard Model fermions, are
charged under this $\Z_4$, and it is then broken along with the
electroweak symmetry when the Higgs acquires its vev. However the
combination $P$ of $R^2$ with a gauge transformation $\Omega$ and a
phase transformation on the fermions remains unbroken even after
electroweak breaking:
\begin{equation}
  P = R^2 \Omega e^{i\frac{\pi}{2} F}, \quad \Omega = 
  \left(
    \begin{array}{c c c}
      -1&0&0\\
      0&-1&0\\
      0&0&1
    \end{array}\right)
\end{equation}
Where $F$ is a fermion number under which $Q,L,\chi$'s and
$\tilde{\chi}$'s have charge $+1$ and $U^c,D^c,E^c,\chi^c$'s and
$\tilde{\chi^c}$'s have charge $-1$. Under P, the Standard Model
Higgses and fermions are even,
\begin{equation}
  h_{1,2} \to h_{1,2}, \quad \psi_{SM} \to \psi_{SM}
\end{equation}
while $\phi,\eta$ are odd 
\begin{equation}
  \phi \to - \phi, \quad \eta \to - \eta
\end{equation}
Since this is an unbroken symmetry even after EWSB, $\phi$ and $\eta$
do not acquire vevs as claimed earlier. Furthermore, the lightest
scalar in the $\phi,\eta$ sector will be exactly stable. Depending on
the parameters of the theory, this stable particle may be neutral, a
linear combination of $\phi_3$ and $\eta$. This  particle
with a mass  in the $\sim 100$ GeV range provides a natural WIMP
candidate for the dark matter of the universe, with weak-scale mass
and cross-section.
 
\section{$N=2$}
In this section we examine the smallest model, with $N=2$, in detail.
The price we pay for using such a small moose is the presence of small
loops in the diagram, which are associated with operators generated
with log divergent coefficients at one-loop. These operators must then be
included in our Lagrangian with arbitrary coefficients of natural size
$\sim g^4/(4 \pi)^2$. For $N\geq 3$, the natural size of the analogous
operators is much smaller, and therefore negligible compared to the
finite one-loop radiative correction.

\begin{figure}
  \centering
  \includegraphics[width=14cm]{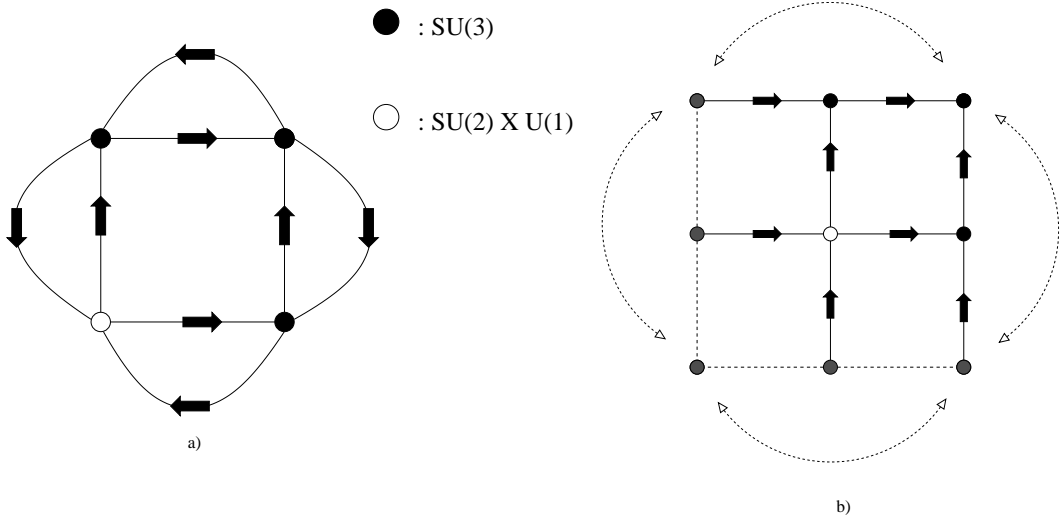}
  \caption{Moose diagrams for the $N=2$ torus. b) is an unfolding of
    a). There are 4 independent sites: in b), any two sites related by
    a reflection through the vertical or the horizontal axis are
    identified. An example of this identification is shown by the
    dotted arrows.  There are 8 independent link fields represented by
    the thick arrows. a) shows clearly the sites and links while b)
    makes the plaquettes and all the discrete symmetries apparent.}
  \label{fig:Neq2moose}
\end{figure}

\begin{figure}[h]
  \centering
  \includegraphics[width=14cm]{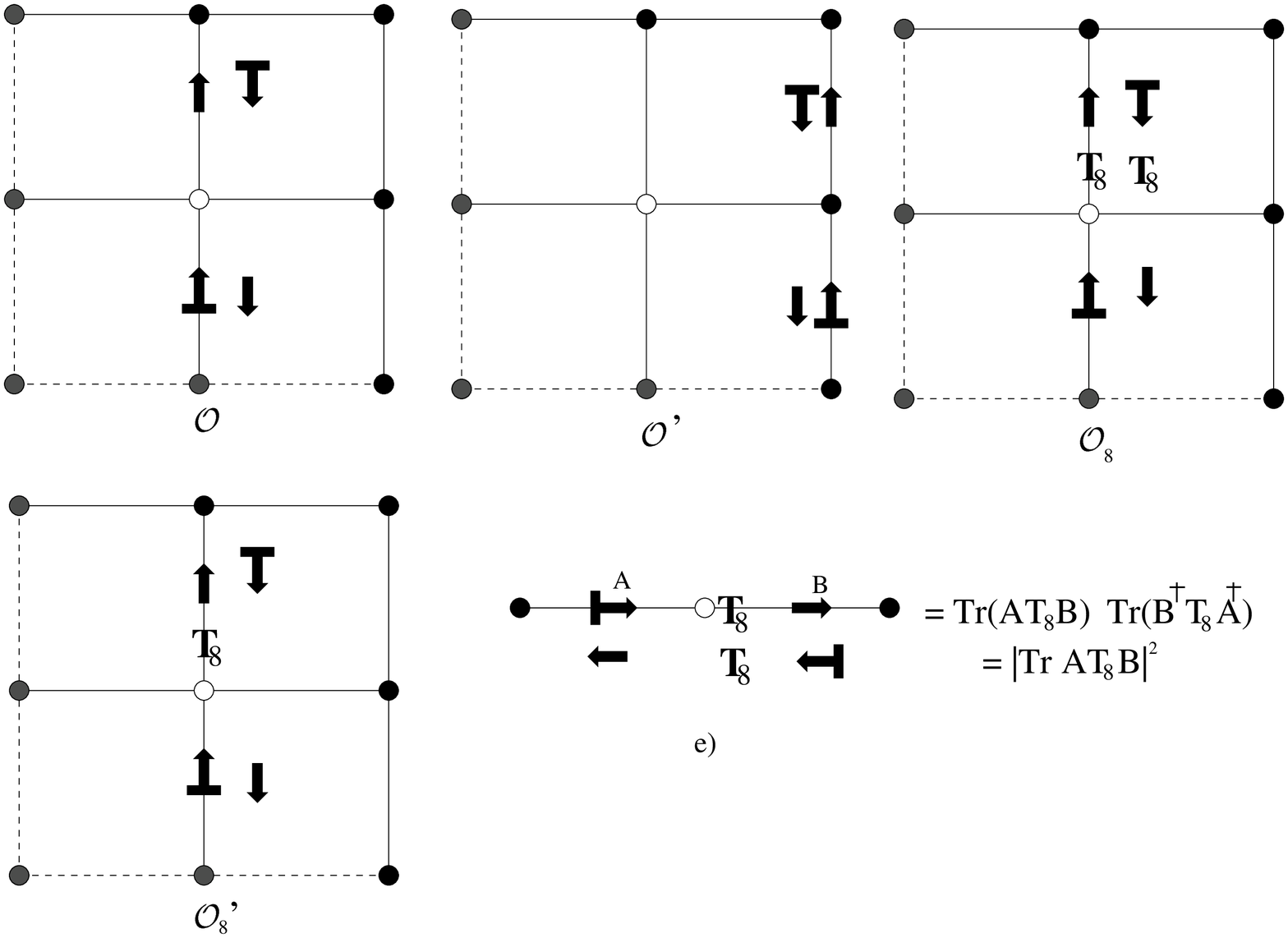}
  \caption{Ooperators ${\cal O}$, ${\cal O}'$,${\cal
      O}_8$ and ${\cal O}_8'$ included in the $N=2$ torus Lagrangian.
    They consist of a product of links forming a closed path. The links
    included in the product are represented by the arrows, and the
    large base indicates a trace should be included before the
    corresponding link. The $T_8$ indicates that a $T_8$ matrix should
    be inserted in the product. The final $\Z_4$ invariant operators
    are obtained by summing over the $90^{\circ}$ rotations of these
    pictures.  e) is an example of this notation relevant for ${\cal
      O}_8$.}
  \label{fig:operator1}
\end{figure}

\begin{figure}[h]
  \centering
  \includegraphics[width=10cm]{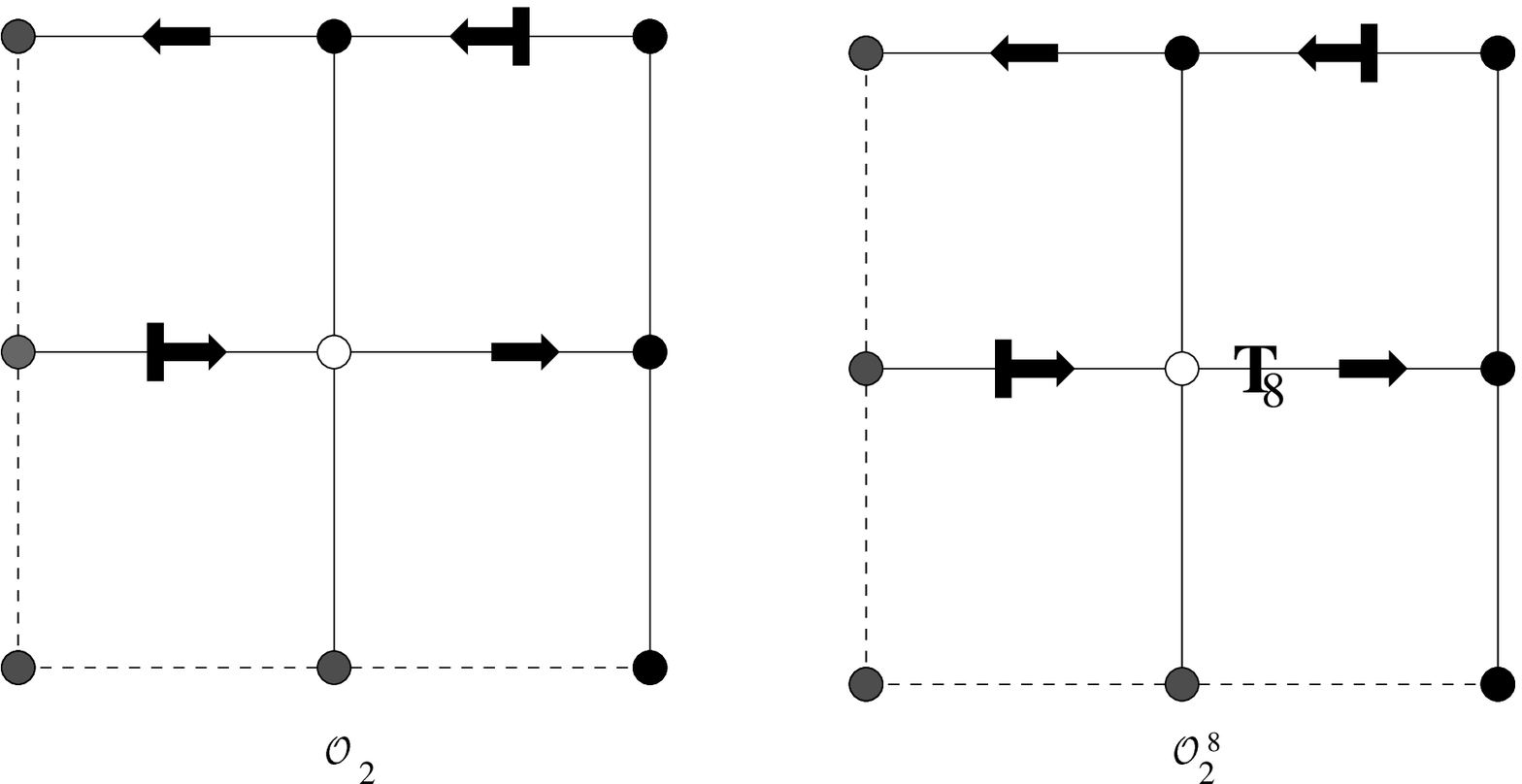}
  \caption{Operators ${\cal O}_2$and  ${\cal O}^8_2$ can be
    generated by the plaquette. Again, they are gauge invariant
    products of the links represented by arrows, the large base
    representing a trace and they are implicitly
    symmetrized  by $90^{\circ}$ rotations.}
  \label{fig:operator2}
\end{figure}

The gauge group of this model is $SU(3)^3\times SU(2) \times U(1)$.
The $\Z_4$ symmetry allows for 2 independent $SU(3)$ couplings, but
for simplicity we will take them to have a common value $g_3$. The
$SU(2)$ and $U(1)$ gauge couplings are $g_2$ and $g_1$. The moose
diagram for the model is shown in Fig. \ref{fig:Neq2moose}. The
relevant parts of the Lagrangian are the kinetic terms, the 4
plaquettes, the 4 $T_8$ plaquettes, the fermion interactions and the
operators generated with log divergence mentioned earlier.
For this particular model, the notation introduced earlier for the
general $N$ torus is incomplete since there are two different links
between every pair of adjacent sites. Therefore, instead of writing
down formul\ae, we will represent these operators by pictures. The
operators that can be generated purely through gauge interactions are
represented in Fig. \ref{fig:operator1}.  We call the resulting operators
${\cal O}$, ${\cal O}'$, ${\cal O}_8$ and ${\cal O}_8'$.
These operators appear with coefficient of natural size $g^4/(4 \pi)^2
f^4$. We also have two operators induced through plaquette
interactions, ${\cal O}_2$ and ${\cal
  O}^8_2$,  shown in Fig. \ref{fig:operator2}.
These operators need two plaquettes to be generated and have natural
size $\sim \lambda^2 /(16 \pi^2) f^4 \sim g^4/(4 \pi)^2 f^4$.  Thus
our Lagrangian contains:
\begin{equation}
  \LL_{\text{other}}= \frac{g_3^4}{16 \pi^2}f^4 \left(a {\cal
      O}_1+a'{\cal O}'_1+a_8{\cal O}_1^8 + a_8' {\cal O}_8' +b {\cal
      O}_2 + b_8 {\cal O}_2^8\right) + \cdots
\end{equation}
where the ellipses represent subdominant terms of order $g^6/(4\pi)^4$
and higher. Note that we have scaled the coefficients of these
operators with the largest gauge coupling; their natural sizes might
somewhat smaller, corresponding to small values of the 
$a$'s and $b$'s.
\subsection{Very low energy theory}
The very low energy theory contains the SM particles, two Higgs
doublets $h_1,h_2$, one complex $SU(2)$ triplet scalar $\phi$ and one
complex singlet scalar $\eta$.  The scalar potential comes from three
sources: first the plaquettes give a quartic interaction for the
little Higgses.  Second, the operators in $\LL_{\text{other}}$ give
mass terms. We assume that these mass squared terms are all positive.
Finally, in order to break the electroweak symmetry, we need a
negative mass squared for the Higgs. As advertised, this is achieved
through the one loop top Yukawa contribution to the Higgs potential
which was given in equation \eqref{eq:topCW}:
\begin{equation*}
  -\left(\frac{M^2 {M_R'}^2 {M_L'}^2}{2 \pi^2 f^2 ({M_R'}^2-{M_L'}^2)}
    \log     \frac{M^2+4 {M_R'}^2}{M^2+4 {M_L'}^2}\right) \abs{h_1}^2
  +\mathcal{O}(h^4)\equiv -\Delta_{m_1}^2\abs{h_1}^2  
\end{equation*}
To quartic order, the potential for the Higgs doublets is then given
by:
\begin{multline}
  V(h_1,h_2) = 2 \left(\lambda - \frac{\alpha}{2}\right)
  \left(\abs{h_1}^4+\abs{h_2}^4\right) -2 
  \left(\lambda+\alpha\right) \abs{h_1^{\dagger}h_2}^2 \\-
  2 \left(\lambda -2 \alpha\right) \abs{h_1}^2 \abs{h_2}^2
  + (m^2 -\Delta_{m_1}^2) \abs{h_1}^2 + m^2 \abs{h_2}^2
\end{multline}
where $m^2$ is given in term of the coefficients appearing in
$\LL_{\text{other}}$:
\begin{equation}
  \label{eq:m}
  m^2 =  \frac{g_3^4 f^2}{16 \pi^2}\left[-24(a+a'+b) + 6
    (b_8+a_8')\right] 
\end{equation}
Assuming $m^2 < \Delta_{m_1}^2$ and $ 2 m^2 -\Delta_{m_1}^2 >0 $,
$h_1$ gets a vev:
\begin{equation}
  h_1 = \begin{pmatrix} 0 \\ v/2\end{pmatrix} \qquad v^2
  = \frac{\Delta_{m_1}^2-m^2}{2\left(\lambda- \frac{\alpha}{2}\right)}
  =\left(246 \text{ GeV}\right)^2 
\end{equation}
The physical spectrum is easily derived in unitary gauge where:
\begin{equation}
  h_1 = \begin{pmatrix} 0 \\ \left(v + h_0\right)/\sqrt{2}
  \end{pmatrix} \quad h_2 = 
  \begin{pmatrix} H^+ \\ \left(H_0+i A_0\right)/\sqrt{2}\end{pmatrix}
\end{equation}
The masses are:
\begin{equation}
  \begin{split}
  m^2_{h_0} =& 2 (\Delta_{m_1}^2-m^2)
  =4\left(\lambda-\frac{\alpha}{2}\right) v^2 \\ 
  m^2_{H_0} =& m^2_{A_0} = 2 m^2 - \Delta_{m_1}^2\\ 
  m^2_{H^+} =& m^2 - \left(\lambda-2 \alpha\right) v^2
  \end{split}
\end{equation}

Note that in this model $H_0$ cannot decay to fermions or gauge
bosons at tree level. After EWSB, the quartic potential generates
cubic terms of the form $H_0 \phi \phi,H_0 \phi \eta,H_0 \eta \eta$
and therefore these decays will dominate if kinematically allowed.
Otherwise $H_0$ will decay at loop level, for instance to gauge
bosons via light scalar loops. Similar considerations apply to $H^+$.
Next we turn to the triplet and singlet masses. There are two
electrically charged scalars, $\phi_1^+ $ and $\phi_2^+$, with masses
\begin{equation}
  \begin{split}
    m_{\phi_1^+}^2 =& -\left(\lambda+\alpha\right) v^2 + m^2 -18
    \frac{g_3^4}{16 \pi^2} (b_8+a_8') f^2 \\
    m_{\phi_2^+}^2 =& \left(2\lambda+\frac{\alpha}{2}\right) v^2 + m^2
    -18 \frac{g_3^4}{16\pi^2} (b_8+a_8') f^2
  \end{split}
\end{equation}
The mass splitting can be related to the Higgs doublets masses:
\begin{equation}
  m_{\phi_1^+}^2 - m_{\phi_2^+}^2 = -\frac{1}{4} m_{h_0}^2 + 2
  \left(m_{H_0}^2 - m_{H^+}^2\right)
\end{equation}
There are also 2 complex neutral scalars, $N_1$ and $N_2$, which are
linear combinations of $\phi_3$ and $\eta$. The lightest of $\phi^+_1$,
$\phi^+_2$, $N_1$ and $N_2$ is stable since  all these are odd under
the discrete $P$ symmetry introduced in section \ref{sec:dark}.
Phenomenologically, we need the neutral states to be lightest for
dark matter. The charged states will then typically decay to neutral
states along with (on or off-shell) $W$'s.

\subsection{Heavy modes}
In addition to the little Higgses, the theory contains heavy TeV scale
modes that are responsible for stabilizing the weak scale: 3 $SU(2)$
triplets, $W_{1,2,3}^a$, 3 $SU(2)$ doublets $V^i_{1,2,3}$
and 3 $SU(2)$ singlet $B_{1,2,3}$ vector bosons ; 1 complex and 1 real
$SU(2)$ triplet $\phi'_1,\phi'_3$, 3 $SU(2)$ doublets $h'_{1,2,3}$ and
1 complex and 1 real $SU(2)$ singlet scalars $\eta'_1,\eta'_3$. There
are also heavy fermions which we denote collectively by $\chi$. Table
\ref{table:heavy} gives the masses of these modes before EWSB, their
quantum numbers, their main decay modes and their transformation
properties under the unbroken $P$ symmetry introduced in section
\ref{sec:dark}. For gauge bosons, the masses are calculated by
diagonalizing the mass matrix found by setting the link fields to
unity in the kinetic term. The scalar mass matrix is extracted from
the plaquette potential, expanding the exponentials to quadratic
order.  Likewise, the interaction between the different modes are
given by expanding the plaquettes and kinetic terms. The kinetic terms
give vector-vector-scalar and scalar-scalar-vector couplings while the
plaquettes give scalar-scalar-scalar coupling. It is straightforward
to see that, when the little Higgses are uniform in theory space,
there are no little Higgs-little Higgs-heavy scalar couplings.

\subsection{Spectrum}
\begin{figure}
\centering
\includegraphics[width=12 cm]{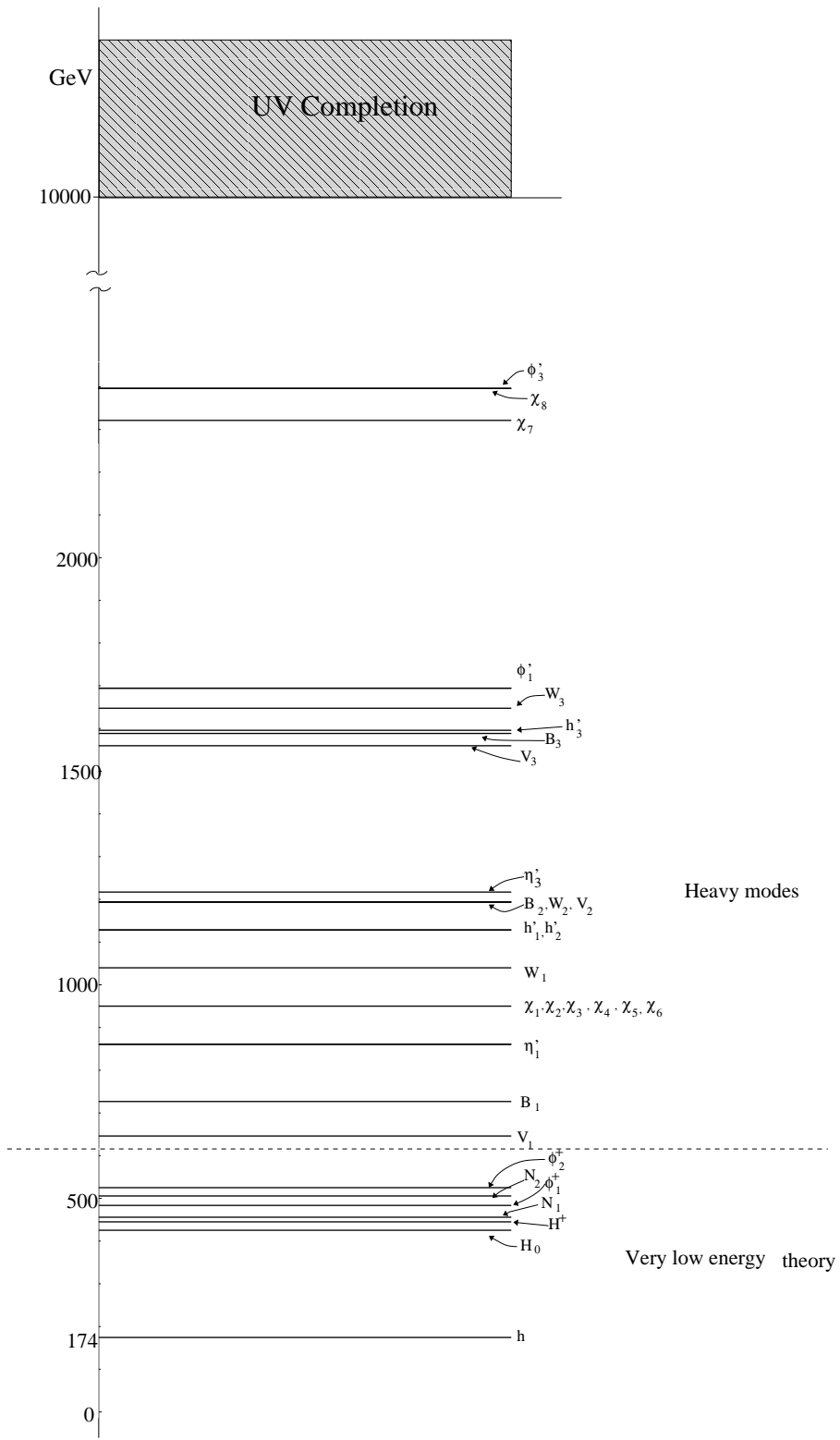}
\caption{Spectrum of the theory.}
\label{fig:spectrum}
\end{figure}

As an illustration, we have chosen a typical set of parameters and
calculated the full spectrum of the theory. For the input parameters in
Table
\ref{table:para}, the resulting spectrum is shown in
Fig. \ref{fig:spectrum}. With this choice of parameters, the lightest Higgs
mass is 174 GeV at tree level, and the UV completion scale is at 10 TeV.   
It is interesting to compare the values of the input
coefficients for the special operators with the low-energy one-loop
log-divergent radiative corrections, ignoring the log factor. 
In all cases, these input values are comparable to the one-loop corrections,
demonstrating the absence of fine-tuning in these couplings. 
Finally, the finite negative contribution to the Higgs mass squared from
the large top Yukawa coupling is $-(460$ GeV$)^2$. For our light Higgs mass
$\sim 174$ GeV, this required a bare positive mass squared of
$+(442$GeV$)^2$. These mass squared choices are moderately tuned at
the 10\% level. This modest tuning can be reduced either with a somewhat
heavier Higgs or with lighter masses for the vector-like $\chi$ fermions.

\begin{table}
\begin{equation*}
\begin{array}{|c|c|c|c|c|c|c|} \hline
\text{particle}&\text{spin}&\text{mass}^2&SU(2)&U(1) & P&\text{decay} \\ \hline
W_{1}^{a}&1&g_3^2 f^2 \beta^-_2&
\mathbf{3} &0&+&f^+ f^-\\ \hline
W_{2}^{a}&1 &g_3^2 f^2& \mathbf{3} &0&+&\phi \phi^{\text{ }\ddagger} \\ \hline
W_{3}^{a}&1&g_3^2 f^2 \beta^+_2&
\mathbf{3} &0&+&f^+ f^-\\ \hline
B_{1}&1&g_3^2 f^2 \beta^-_1& \mathbf{1} &0
&+&f^+ f^-\\ \hline
B_{2}&1& g_3^2 f^2& \mathbf{1} &0 &+&h h^{\text{ }\ddagger}\\ \hline
B_{3}&1& g_3^2 f^2 \beta^+_1& \mathbf{1} &0
&+&f^+ f^-\\ \hline
V_1^i& 1& g_3^2 f^2(1-\sqrt{2}/2)& \mathbf{2}&1/2&
-&\phi^{\dagger} h\\ \hline
V_2^i& 1& g_3^2 f^2& \mathbf{2}&1/2& -&\phi h^{\text{ }\ddagger}\\ \hline
V_3^i& 1& g_3^2 f^2 (1+\sqrt{2}/2)& \mathbf{2}&1/2&
-&\phi^{\dagger} h\\ \hline
h'_1&0 &16 \left(\lambda-\frac{\alpha}{2}\right) f^2 &
\mathbf{2}& 1/2&+&t \bar{t},t\bar{b}\\ \hline
h'_2&0 &16 \left(\lambda-\frac{\alpha}{2}\right) f^2 &
\mathbf{2}& 1/2&+&h_1^{\dagger} \phi_3',h_1^{\dagger} \eta'_3,h_1^{\dagger}
W_2, \phi^{\dagger}V_2,h_2^{\dagger} W_{1,3},t \chi_1^c \\ \hline
\phi'_1&0 &16 \left(\lambda+\alpha\right) f^2 &
\mathbf{3}& 0&-&{h_3'}^{\dagger} h_{1},\phi W_2,\phi^{\dagger} W_{1,3}, t\chi_{5,}^c \\ \hline
\eta'_1&0 &16 \left(\lambda-\alpha\right) f^2 &
\mathbf{1}& 0&-&h_2 V_2, h_2^{\dagger} V_{1,3}, t \chi_5^c \\ \hline
h'_3&0 &32 \left(\lambda-\frac{\alpha}{2}\right) f^2 &
\mathbf{2}& 1/2&-&h_2^{\dagger} \phi_1',W V_2^{\text{ }\ddagger},t \chi_3^c \\ \hline
\phi'_3&0 &32 \left(\lambda+\alpha\right) f^2 & \mathbf{3}&
0&+&{h'}^{\dagger}_1 h_2, t\chi_4^c,W W_2^{\text{ }\ddagger} \\ \hline
\eta'_3&0 &32 \left(\lambda-\alpha\right) f^2 & \mathbf{1}&
0&+&h_{1}^{\dagger}h_2',t \chi_4^c \\ \hline
\chi_1&1/2&M^2&\mathbf{1}& \pm 2/3&+& h_2 t,  h_2 b \\ \hline
\chi_2&1/2&M^2&\mathbf{1}& \pm 2/3&-&\eta t^c \\ \hline
\chi_3&1/2&M^2&\mathbf{1}& \pm 2/3&-&\eta t^c \\ \hline
\chi_4&1/2&M^2&\mathbf{2}&\pm 1/6&+&h_2 t^c \\ \hline
\chi_5&1/2&M^2&\mathbf{2}&\pm 1/6&-&\phi t,\eta t,\phi b,\eta b \\ \hline
\chi_6&1/2&M^2&\mathbf{2}&\pm 1/6&-&\phi^{\dagger} t,\eta^{\dagger}
t,\phi^{\dagger}b,\eta^{\dagger} b \\ \hline
\chi_7&1/2&M^2+4 {M_L'}^2&\mathbf{2}&\pm1/6&+&h_1 t^c\\ \hline
\chi_8&1/2&M^2 + 4 {M_R'}^2&\mathbf{1}&\pm2/3&+&h_1 t, h_1 b\\ \hline
\end{array}
\end{equation*}
\caption{Summary of heavy modes. Here $\beta^\pm_i =
1/2\left(2+g_i/g_3\pm\sqrt{2-2g_i^2/g_3^2+g_i^4/g_3^4}\right)$. The decay
mode with the ${}^{\text{ }\ddagger}$ symbol vanish when all the $f$'s are equal.}
\label{table:heavy}
\end{table}

\begin{table}
\begin{equation*}
\begin{array}{|c|c|c|c|} \hline
\text{parameter} & \text{input value}&\text{1 loop vector}&\text{1 loop
scalars}\\ \hline
\Lambda & 10 \text{ TeV}&& \\ \hline
f & 795 \text{ GeV} &&\\ \hline
g_1 & 0.21& & \\ \hline
g_2 & 0.995& & \\ \hline
g_3 & 1.5 & &\\ \hline
M & 950 \text{ GeV} & &\\ \hline
M_R'&1100 \text{ GeV} & &\\
\hline
\lambda & 0.18 & &\\ \hline
a & 1.06 & &\\ \hline
a' & 0.3 & &\\ \hline
b & -1.8 & &\\ \hline
a_8'&-0.08 &-0.034&0\\ \hline
a_8 & 0.04 &-0.0052&0\\ \hline
b_8 & -0.07 &0&0.135y\\ \hline\hline
m^2 &\left(442 \text{GeV}\right)^2& \left(226 \text{GeV}\right)^2&\left(202
\text{GeV}\right)^2
\\ \hline
\end{array}
\end{equation*}
\caption{Input parameter used to generate the spectrum. $M_L'$ is then
fixed by the measured top Yukawa coupling to be $M_L' = 1060$ GeV, and 
$\alpha$ is fixed by the
scale of EWSB $v = 246$ GeV to be $\alpha = 0.1$. Note that $m^2$ is
given as a function of the other parameters (see eq \eqref{eq:m})}
\label{table:para}
\end{table}

\section{Conclusion}

We have studied a class of realistic models of electroweak symmetry
breaking in which the Higgs appears as an extended object in theory
space. This class of models, first introduced in
\cite{Arkani-Hamed:2001nc}, is based on toroidal theory spaces. We
reviewed and extended the analysis of the $N \times N$ torus and
studied the $N=2$ case in detail.

Although our analysis was based on a specific theory space, the
mechanism of EWSB applies more generally \footnote{The general rules
  for building such models, together with the procedure for
  identifying the little Higgses from topological properties of the
  theory space, will be discussed in \cite{moose}.}.  A robust
prediction of all such models is the presence of at least two
naturally light Higgs doublets. This is a consequence of the Higgs
quartic couplings descending from plaquette interactions, producing
commutator squared potentials. In the specific models studied in this
paper there are additional weak triplets and singlets scalar at low
energy.  Theory spaces often have geometric discrete symmetries, such
as the $\Z_4$ symmetry which is broken to a $\Z_2$ after electroweak
symmetry breaking in the model we discussed. The light triplets and
singlets are odd under this symmetry and therefore the lightest of
these states is stable, providing a WIMP dark matter candidate. We
also showed that the top quark can drive the Higgs mass negative and
trigger EWSB in a natural way.  At TeV energies, new physics appears
to stabilize the weak scale. This new physics takes the form of new
heavy particles which soften each of the quadratic divergences of the
low energy theory: heavy gauge bosons cancel quadratically divergent
gauge boson loops, heavy scalars cancel the quadratic divergence from
the low energy quartic couplings and heavy fermions cancel the
quadratic divergence from the top loop.

In the $N=2$ case, we computed the full spectrum of the theory for a
typical set of parameters, and briefly discussed the primary decay
modes for the new fields, leaving a more detailed study of both the
collider phenomenology and cosmological implications of this theory to
future work.

How do our theories compare with supersymmetric models of electroweak
symmetry breaking? One of the features that makes supersymmetry
attractive is its ability to accommodate precision electroweak data,
as a consequence of its weakly coupled description of EWSB. Since our
models are weakly coupled at the TeV scale, they share this success.
In supersymmetry, there is often a flavor problem associated with
flavor-changing neutral currents for the light generations, mediated
by superpartners.  In our models, the only new states which carry
flavor are the heavy fermions which ``delocalize'' the top quark in
theory space, and so there are no flavor-changing interactions
involving the first two generations at the TeV scale. (This is similar
to the situation in the ``more minimal'' supersymmetric models with
heavy first two generation scalars \cite{Cohen:1996vb,Dimopoulos:1995mi}).

The supersymmetric Standard Model also has the virtue that it can be
extrapolated to energies much higher than the TeV scale.  A
remarkable success of such an extrapolation is the supersymmetric
prediction of gauge coupling unification. This prediction remains
one of the most compelling arguments for  supersymmetry and
the energy desert.

By contrast, our effective theory description of the physics breaks
down at relatively low scales, $\sim$ 10--100 TeV. There are
straightforward UV completions of our non-linear sigma models into
linear sigma models, but like the Standard Model, these themselves
suffer from a (slightly reduced) ``hierarchy problem''. We can instead
consider supersymmetrizations of these linear sigma models above $\sim$
10--100 TeV, or perhaps strongly coupled gauge dynamics or even low
scale gravity models above this scale.  Whatever this UV completion
physics is, it completely decouples from the physics of EWSB, and is
therefore irrelevant for upcoming collider experiments.  As for gauge
coupling unification, there have been a number of recent proposals on
retaining this success at low energies (see for instance
\cite{Arkani-Hamed:2001vr,Dienes:1998vh}), or recovering the correct
prediction of the weak mixing angle from different group-theoretic
structures \cite{Weinberg:1972nd,Dimopoulos:2002mv}. It would be
interesting to pursue combining these ideas with our models, and some
work along these lines is already in progress \cite{dimkap}.

\section{Acknowledgements}
N.A-H. and A.G.C. would like to thank Howard Georgi for valuable comments
 and J.G.W. and T.G. would like to thank Takemichi Okui for useful
discussions. J.G.W. and T.G. thank the Harvard theory group for the use of
their facilities. A.G.C. is supported in part by the Department of Energy under
grant number \# DE-FG02-91ER-40676. N.A-H. is supported in part by the
Department of Energy under Contracts DE-AC03-76SF00098, The National
Science Foundation under grant PHY-95-14797, The Alfred P. Sloan
foundation, and the David and Lucille Packard Foundation. T.G. is also supported by an NSERC fellowship.   


\begin{thebibliography}{10}

\bibitem{Arkani-Hamed:2001nc}
N.~Arkani-Hamed, A.~G. Cohen, and H.~Georgi, ``Electroweak symmetry breaking
  from dimensional deconstruction,'' {\em Phys. Lett.} {\bf B513} (2001)
  232--240, \href{http://xxx.lanl.gov/abs/hep-ph/0105239}{{\tt
  hep-ph/0105239}}.

\bibitem{Arkani-Hamed:2001ca}
N.~Arkani-Hamed, A.~G. Cohen, and H.~Georgi, ``(De)constructing dimensions,''
  {\em Phys. Rev. Lett.} {\bf 86} (2001) 4757--4761,
  \href{http://xxx.lanl.gov/abs/hep-th/0104005}{{\tt hep-th/0104005}}.

\bibitem{Hill:2000mu}
C.~T.~Hill, S.~Pokorski and J.~Wang,
Phys.\ Rev.\ D {\bf 64}, 105005 (2001)
[arXiv:hep-th/0104035].

\bibitem{Kaplan:1984fs}
D.~B. Kaplan and H.~Georgi, ``SU(2) x U(1) breaking by vacuum misalignment,''
  {\em Phys. Lett.} {\bf B136} (1984) 183.

\bibitem{Kaplan:1984sm}
D.~B. Kaplan, H.~Georgi, and S.~Dimopoulos, ``COMPOSITE HIGGS SCALARS,'' {\em
  Phys. Lett.} {\bf B136} (1984) 187.

\bibitem{Weinberg:1979kz}
S.~Weinberg, ``Phenomenological Lagrangians,'' {\em Physica} {\bf A96} (1979)
  327.

\bibitem{Manohar:1984md}
A.~Manohar and H.~Georgi, ``Chiral quarks and the non-relativistic quark
  model,'' {\em Nucl. Phys.} {\bf B234} (1984) 189.

\bibitem{moose}
N.~Arkani-Hamed, A.~G. Cohen, T.~Gregoire, and J.~Wacker, ``Mooses, Topology
  and Higgs.'' in preparation.

\bibitem{Cohen:1996vb}
A.~G. Cohen, D.~B. Kaplan, and A.~E. Nelson, ``The more minimal supersymmetric
  standard model,'' {\em Phys. Lett.} {\bf B388} (1996) 588--598,
  \href{http://xxx.lanl.gov/abs/http://arXiv.org/abs/hep-ph/9607394}{{\tt
  http://arXiv.org/abs/hep-ph/9607394}}.

\bibitem{Dimopoulos:1995mi}
S.~Dimopoulos and G.~F. Giudice, ``Naturalness constraints in supersymmetric
  theories with nonuniversal soft terms,'' {\em Phys. Lett.} {\bf B357} (1995)
  573--578,
  \href{http://xxx.lanl.gov/abs/http://arXiv.org/abs/hep-ph/9507282}{{\tt
  http://arXiv.org/abs/hep-ph/9507282}}.

\bibitem{Arkani-Hamed:2001vr}
N.~Arkani-Hamed, A.~G. Cohen, and H.~Georgi, ``Accelerated unification,''
  \href{http://xxx.lanl.gov/abs/hep-th/0108089}{{\tt hep-th/0108089}}.

\bibitem{Dienes:1998vh}
K.~R. Dienes, E.~Dudas, and T.~Gherghetta, ``Extra spacetime dimensions and
  unification,'' {\em Phys. Lett.} {\bf B436} (1998) 55--65,
  \href{http://xxx.lanl.gov/abs/hep-ph/9803466}{{\tt hep-ph/9803466}}.

\bibitem{Weinberg:1972nd}
S.~Weinberg, ``Mixing angle in renormalizable theories of weak and
  electromagnetic interactions,'' {\em Phys. Rev.} {\bf D5} (1972) 1962--1967.

\bibitem{Dimopoulos:2002mv}
S.~Dimopoulos and D.~E. Kaplan, ``The weak mixing angle from an SU(3) symmetry
  at a TeV,''
  \href{http://xxx.lanl.gov/abs/http://arXiv.org/abs/hep-ph/0201148}{{\tt
  http://arXiv.org/abs/hep-ph/0201148}}.

\bibitem{dimkap}
D.~E. Kaplan and S.~Dimopooulos,  in preparation.


\end{thebibliography}

\providecommand{\href}[2]{#2}\begingroup\raggedright\endgroup

\end{document}